\documentclass{aa}  
\usepackage{graphicx}
\usepackage{txfonts}
 \usepackage{graphicx}
 \usepackage[normalem]{ulem}
 \useunder{\uline}{\ul}{}
 \usepackage{lscape}
 \usepackage{xcolor}
 
\begin{document} 

\title{Revisiting empirical solar energetic particle scaling relations}

  \subtitle{II. Coronal mass ejections}

   \author{Athanasios Papaioannou
          \inst{1}
          \and
          Konstantin Herbst\inst{2}
        \and
          Tobias Ramm \inst{\textbf{3}}
        \and
          David Lario \inst{4}
       \and
          Astrid M. Veronig \inst{5}
   }       

   \institute{Institute for Astronomy, Astrophysics, Space Applications and Remote Sensing (IAASARS), National Observatory of Athens, I. Metaxa \& Vas. Pavlou St., 15236 Penteli, Greece\\
              \email{atpapaio@astro.noa.gr}
         \and
             Institut f\"ur Experimentelle und Angewandte Physik, Christian-Albrechts-Universit\"at zu Kiel, 24118 Kiel, Germany
             \and
        Institut f\"ur Theoretische Physik und Astrophysik, Christian-Albrechts-Universit\"at zu Kiel, 24118 Kiel, Germany
       \and
        NASA, Goddard Space Flight Center, Heliophysics Science Division, Greenbelt, MD 20771, USA
        \and
       Institute of Physics \& Kanzelh\"{o}he Observatory for Solar and Environmental Research, University of Graz, A-8010 Graz, Austria}

  \titlerunning{Solar scaling relations. Part II}  

 
  \abstract
   {}
   {The space radiation environment conditions and the maximum expected coronal mass ejection (CME) speed are assessed by investigating scaling laws between the peak proton flux and fluence of solar energetic particle (SEP) events with the speed of the CMEs.}
   {We used a complete catalog of SEP events, covering the last $\sim$25 years of CME observations (i.e., 1997 to 2017). We calculated the peak proton fluxes and integrated event fluences for events that reached an integral energy of up to E$>$ 100 MeV. For a sample of 38 strong SEP events, we first investigated the statistical relations between the recorded peak proton fluxes ($I_{P}$) and fluences ($F_{P}$) at a set of integral energies of E $>$10~MeV, E$>$30~MeV, E$>$60~MeV, and E$>$100 MeV versus the projected CME speed near the Sun ($V_{CME}$) obtained by the Solar and Heliospheric Observatory/Large Angle and Spectrometric Coronagraph (SOHO/LASCO). Based on the inferred relations, we further calculated the integrated energy dependence of both $I_{P}$ and $F_{P}$, assuming that they follow an inverse power law with respect to energy. By making use of simple physical assumptions, we combined our derived scaling laws to estimate the upper limits for $V_{CME}$, $I_{P}$, and $F_{P}$ by focusing on two cases of known extreme SEP events that occurred on 23 February 1956, (GLE05) and in AD774/775, respectively. Based on the physical constraints and assumptions, several options for the upper limit $V_{CME}$ associated with these events were investigated. }
   {A scaling law relating $I_{P}$ and $F_{P}$ to the CME speed as $V_{CME}^5$ for CMEs ranging between $\sim$3400-5400 km/s is consistent with values of $F_{P}$ inferred for the cosmogenic nuclide event of AD774/775. At the same time, the upper CME speed that the current Sun can provide possibly falls within an upper limit of $V_{CME}\leq$ 5500 km/s.}
   {}


   \maketitle
%

\section{Introduction}
\label{sec:intro}
Solar energetic particle (SEP) events result from acceleration processes associated with both solar flares and coronal mass ejections (CMEs). SEPs propagate in interplanetary space mostly along interplanetary magnetic field (IMF) lines before they are observed by spacecraft located in the heliosphere \citep[see][and references there in]{desai2016large,2021LNP...978.....R}. A two-class paradigm classifies the SEP events into impulsive or gradual events. The impulsive SEP events are assumed to be associated with solar flares and type III radio bursts. They are limited in duration, reach low peak intensities, and typically have narrow emission cones \citep[see, e.g.,][]{2021LNP...978.....R}. Gradual SEP events are most energetic and are assumed to be associated with CMEs and type II radio bursts. They can last for several days, achieving elevated peak fluxes, and in general have a broad emission cone \citep[see, e.g.,][]{desai2016large}. Nonetheless, this paradigm has proven to be a simplified view, and it is further complicated by the association of both strong flares and CMEs with SEPs \citep{2010JGRA..115.8101C,2016JSWSC...6A..42P}.

The fact that SEPs are driven by CMEs was first discussed by \cite{1978SoPh...57..429K}, who also indicated the close relation of the CME speed and the peak proton flux of SEPs by highlighting the fact that fast CMEs are more likely to drive shocks that are capable of accelerating energetic particles \citep[see also][]{1982JGR....87.3439K, 2001JGR...10620947K}. This correlation was widely investigated and verified since then \citep[e.g.,][]{2002ApJ...572L.103G, 2010JGRA..115.8101C, 2014SoPh..289.3059R, 2016JSWSC...6A..42P,2017JSWSC...7A..14P,2020ApJ...900...75K}. The interpretation is that shocks driven by fast CMEs are more likely to efficiently accelerate particles since theoretically, the acceleration rate depends on the speed of the shock with respect to the upstream medium \citep{2012SSRv..173..247L}.   

Routine CME observations by the Large Angle and Spectrometric Coronagraph \citep[LASCO;][]{1995SoPh..162..357B} on board the Solar and Heliospheric Observatory (SOHO) have been performed since 1997. These observations have revealed that there is no one-to-one correspondence between X-ray flares and CMEs because many more flares than CMEs are observed. However, most CMEs are associated with some level of X-ray flare emission. In particular, \cite{2006ApJ...650L.143Y} (their Fig. 1) demonstrated that the stronger the flare in terms of its peak flux in the 1--8 $\AA$ soft X-ray (SXR) wavelength band as measured routinely by the GOES satellites, the more likely it is to be associated with a CME. This results from that fact that solar eruptive events (i.e., flares and CMEs) do not occur in isolation, but in concert, as a consequence of corresponding changes in the coronal magnetic field. It is noteworthy that recently, \cite{2021ApJ...917L..29L} reported a critical study of the flare--CME relation, showing that it depends on the flare class and the size of the active region (AR) of the source. Moreover, the solar origin of SEPs can almost always \citep[$\sim$94\%, ][]{2016JSWSC...6A..42P} be associated with the occurrence of both SXR flares and CMEs. Some additional SEP events generated at (or even beyond) the west limb of the Sun \citep{2010JGRA..115.8101C} could only be associated with CMEs because flare observations are not possible when they occur on the far side of the Sun. As a result, a wealth of statistical studies indicate that SXR flare peak fluxes, near-Sun CME speeds, the achieved peak proton intensities, and the fluences of the resulting SEP events are related. From recent studies investigating these correlations, it was shown that the most prominent correlation is found between the SEP peak proton flux and the speed of the CME (i.e., $cc$= 0.57 for E$>$ 10 MeV), with a tendency to decrease at higher particle energies \citep[i.e., 0.40 for E$>$ 100 MeV, see, e.g.][]{2016JSWSC...6A..42P}. 

Scaling relations of the peak proton flux of SEPs ($I_{P}$) to the speed of the CME ($V_{CME}$) have been proposed by several authors. In particular, \cite{2001JGR...10620947K} showed that the relation of the peak proton flux of SEPs at 2 and 20 MeV depends on the $V_{CME}$ in the form $I_{P} \propto V_{CME}^{4.36}$ and  $I_{P} \propto V_{CME}^{4.83}$, respectively. Investigating 130 SEP events at an integral energy of E$>$10 MeV and 88 SEP events at an integral energy of E$>$100 MeV associated with CMEs originating at western longitudes (i.e., W20–W87$^{\circ}$), \cite{2017Ge&Ae..57..727B} found dependences described as $I_{P} \propto V_{n}^{4.02\pm0.39}$ and  $I_{P} \propto V_{n}^{3.01\pm0.50}$, respectively, with $V_{n} = \frac{V_{CME}}{1000}$\footnote{The normalization employed in that paper does not affect the proportionality}. \cite{2014JGRA..119.4185L} showed that the SEP peak intensity versus the CME speed in a good approximation follows a triangular distribution. This method directly provided upper limits to the peak proton particle intensity that can be observed in the prompt component of the SEP events. In particular, these authors showed that for three energy ranges spanning from 9-80 MeV, the resulting upper limit dependence scales with $I_{P} \propto V_{CME}^{\gamma}$ with $\gamma$ ranging from 4.90-5.63 (see their Fig. 5). \citet{2016ApJ...833L...8T} showed on theoretical grounds that the upper limit for the peak proton flux of E$>$10 MeV is proportional to the CME speed as $I_{P} \propto ~V_{CME}^{5}$.  

In our previous study \citep[][ -- hereafter part I]{2023A&A...671A..66P}, we presented the dependence of SEPs on flare parameters. In particular, we investigated the relation between the GOES 1--8 $\AA$ SXR peak flux ($F_{SXR}$) of the parent flare versus the SEP event peak proton flux ($I_{P}$) and fluence ($F_{P}$). We showed that a direct estimation of the upper limit SEP fluence spectra based on $F_{SXR}$ alone is possible and leads to a quantification of the radiation environment. The present follow-up study deals with the dependence of SEPs on the properties of the associated CMEs. With this purpose, we use the catalog presented in detail in part I (Appendix C) and start with the analysis of the relations between $V_{CME}$ and $I_{P}$, and we consider the dependence of $V_{CME}$ on $F_{SXR}$. Based on the findings of \citet{2016ApJ...833L...8T}, showing that the upper limit for the peak proton flux of E$>$10 MeV depends on the CME speed as $I_{P} \propto ~V_{CME}^{5}$, we derive upper limits and scaling relations among the CME speeds ($V_{CME}$) and the achieved SEP peak flux ($I_{P}$) at each integral energy (from E$>$10 to E$>$100 MeV). We then extend these relations to incorporate the fluence ($F_{P}$) of SEPs. Additionally, we deduce the upper-limit fluence spectra of SEPs based on $V_{CME}$, whereas in part I, the SEP fluence spectra were obtained based on $F_{SXR}$. 

In an attempt to estimate upper limits of the extreme events that can stem from our host star, \cite{10.1007/978-3-642-02859-5_24} presented calculations based on a hypothetical AR with the largest reported AR area (i.e., 5000 millionths of a solar hemisphere, msh) and the maximum measured sunspot magnetic field ($B$ = 6100 G). These authors then estimated the potential energy of the AR to be 10$^{36}$ erg, which could produce an SXR flare of $\sim$ X1000 class (i.e., 10$^{-1} W/m^{2}$). Consequently, the estimated maximum CME speed associated with the highest solar flare class, taking into account an upper limit of 26\% of the potential energy being converted into CME kinetic energy, was found to be 7200 km s$^{-1}$. This exceeds the highest measured CME speed by the SOHO/LASCO by a factor of $\sim$ 2 \citep{gopalswamy2011coronal}. By considering the observations used in this work (Sect. \ref{sec:data}), the obtained scaling relations (Sect. \ref{sec:scaling}), and the upper limits our Sun can produce (Sect. \ref{sec:4}), we estimate the fastest expected CMEs, the worst-case SEP proton fluxes and fluences, and the corresponding SEP spectrum. Implications for the effects of extreme CMEs on the radiation environment and the limits of recent fluence reconstructions on $V_{CME}$ are described and discussed.


\section{Data sets}
\label{sec:data}
We focus on the relations between SEPs and CMEs. We used a well-defined catalog of 65 well-connected (W20-90$^{\circ}$) SEP events that were recorded between 1984 and 2017 and extended from E$>$10 to E$>$100 MeV. For each event, we first identified the prompt peak intensity (in units of protons $cm^{-2} sr^{-1} s^{-1}$), defined as the maximum intensity observed shortly after the onset of the event in situ. In this way, energetic storm particles (or the ESP component) were excluded. Furthermore, we calculated the omnidirectional fluence ($cm^{-2}$) (integration over time) and tabulated all results (see Appendix C in part I).  Because SOHO/LASCO  measured linear CME speeds only starting in 1997, the sample used in this work was reduced from 65 to 38 SEP events. The CME speeds and widths were taken from the online CDAW CME catalog\footnote{\url{https://cdaw.gsfc.nasa.gov/CME_list/}} \citep{2004JGRA..109.7105Y}.

\begin{figure}[t!]
\centering
\includegraphics[width=\columnwidth]{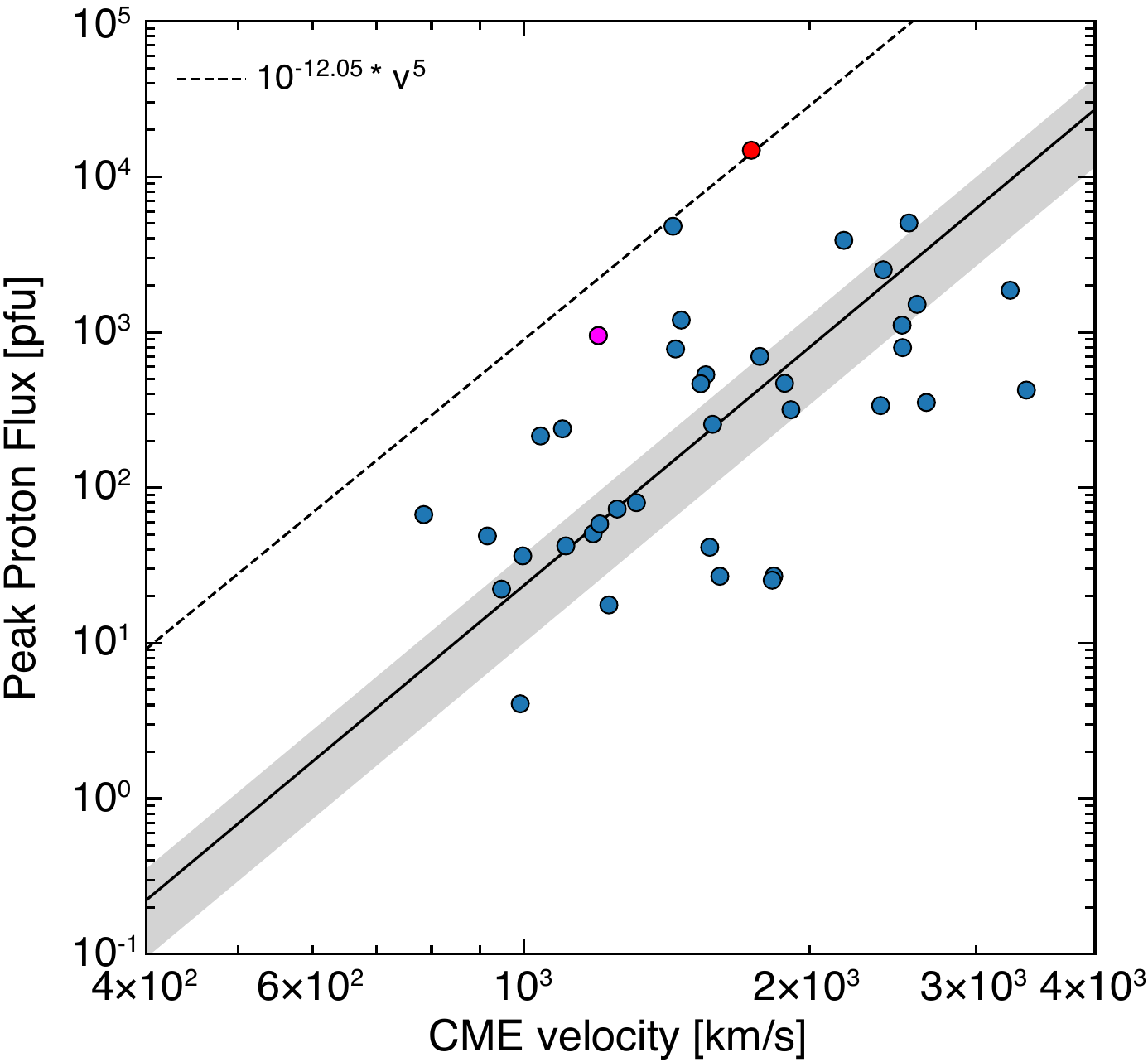}
\caption{$I_P \propto V_{CME}^\gamma$ relation for E$>$ 10 MeV, with $\gamma$=5.09$\pm$0.78. The dashed black line corresponds to the upper limit of $I_P$ in terms of $V_{CME}$ with $\gamma$=5. The upper point in our sample corresponds to the SEP event on 8 November 2000 (see the relevant appendix in part I) and is used for the scaling. It is labeled with a red dot. In addition, the event on 15 April 2001 is highlighted with a magenta dot (see details in the text).} 
\label{fig:fig1}
\end{figure}

\section{Scaling relations}
\label{sec:scaling}
\subsection{Soft X-ray flare flux, coronal mass ejection speed, and peak proton fluxes}
\label{sec:1}
As a first step, we studied the scaling relations between the CME speed ($V_{CME}$) and the prompt peak proton flux ($I_{P}$) for integral energies E$>$ 10 MeV, E$>$ 30 MeV, E$>$ 60 MeV, and E$>$ 100 MeV.  Figure \ref{fig:fig1} shows $I_{P}$ for E$>$10 MeV as a function of $V_{CME}$. The correlation coefficient between $I_{P}$ and $V_{CME}$ for the SEP events that reach an integral energy of E$>$ 10 MeV is $cc$=0.58, assuming a linear regression obtained from the reduced major axis \citep[RMA, see discussion in][]{2023A&A...671A..66P} method, leading to $I_{P} \propto V_{CME}^{\gamma}$, with $\gamma$=5.09$\pm$0.78. Additionally, the gray shaded envelope in Fig.~\ref{fig:fig1} provides the estimated error\footnote{see \citet{2023A&A...671A..66P} for details of the error estimation}. 

As discussed in \citet{2016ApJ...833L...8T}, an upper limit for this relation is given by $I_{P} \propto V_{CME}^{5}$, when passing through the uppermost point of the employed sample. This  scaling relation is in general deduced by employing three assumptions: (a) that the CME mass ($M_{CME}$) is equal to the sum of the gravitational stratified AR corona (see Eq. (1) of \cite{2016ApJ...833L...8T}), (b) that the CME kinetic energy ($E_{CME}$) is proportional to the total energy released during a flare, which is also a fraction $f$ of the AR magnetic field energy \citep[see e.g.][]{2012ApJ...759...71E, 2023A&A...671A..66P} \citep[see Eq.~(2) of][]{2016ApJ...833L...8T}, and (c) that the total kinetic energy of SEPs ($E_{p}$) is proportional to the flare energy, and the duration of the proton flux enhancement is determined by the CME propagation timescale $t_{CME} \propto L/V_{CME}$ ($L$ is the length scale of the flaring AR), which leads to a scaling relation of $ I_{P} \propto V_{CME}^{5}$. Based on our sample, we derived that this upper limit lies at $I_{P} = 10^{-12.05} \cdot V_{CME}^{5}$ (dashed black line in Fig.~\ref{fig:fig1}). The event that marks the upper limit fit is the event on 8 November 2000 SEP (indicated by a red dot in Fig. \ref{fig:fig1}). We further highlighted the ground-level enhancement event of 15 April 2001 (GLE60) in magenta (see details in Appendix \ref{appendix:B}).  

\begin{figure}[!t]
\centering
\includegraphics[width=\columnwidth]{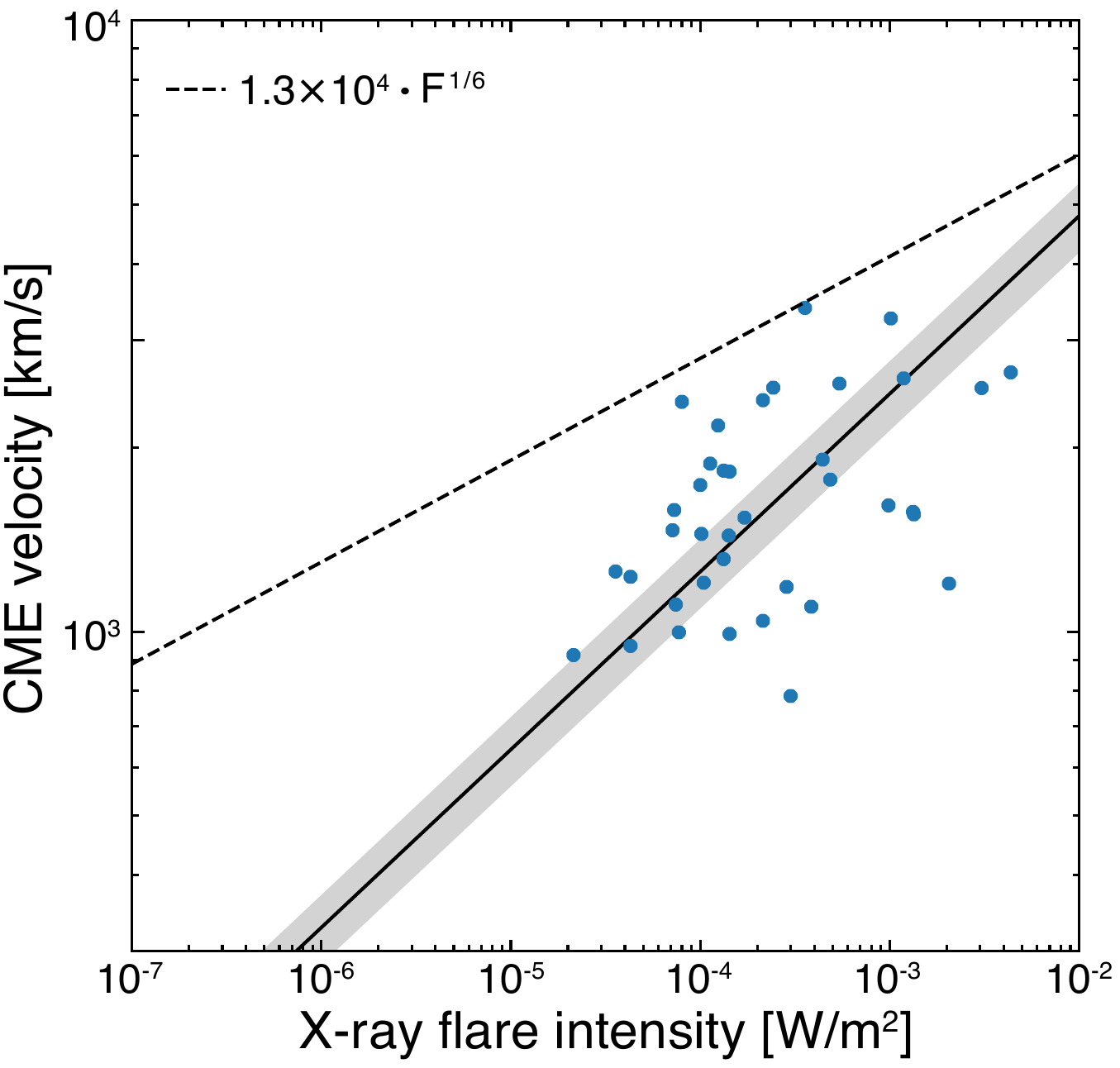}
\caption{Relation between CME velocity ($V_{CME}$) and the flare SXR peak flux ($F_{SXR}$). The solid black line represents the linear RMA regression fit $V_{CME}$ $\propto$ $F_{SXR}^{\alpha}$ with $\alpha$ = 0.43$\pm$0.08. The dashed black line corresponds to the upper limit of $V_{CME}$ in terms of $F_{SXR}$. See text for details.} 
\label{fig:fig2}
\end{figure}

A scatter plot between $V_{CME}$ and the flare SXR peak flux ($F_{SXR}$) is presented in Fig.~\ref{fig:fig2}. The RMA regression is quantified as $V_{CME}$ $\propto$ $F_{SXR}^{0.43\pm 0.08}$ with a correlation coefficient of $cc$=0.43. \citet{2016ApJ...833L...8T} suggested that the upper limit of this relation is given by $V_{CME}$ $\propto$ $F_{SXR}^{1/6}$. Following the description by \cite{2016ApJ...833L...8T}, based on our sample, we find the relation to be $V_{CME}$ = $1.3 \times 10^{4} \cdot F_{SXR}^{1/6}$ (dashed black line in Fig. \ref{fig:fig2}), passing the point with the highest $V_{CME}$ = 3387 km/s corresponding to the event on 10 November 2004.

The results presented to date assumed statistical relations among the SXR peak flux of flares ($F_{SXR}$), the speed of the CME ($V_{CME}$), and the peak proton flux ($I_{P}$) at an integral energy of E$>$10 MeV \citep[e.g.][]{2016ApJ...833L...8T}. Solar flares and CMEs are both drivers of SEP events \citep[e.g.][]{2010JGRA..115.8101C}, and a correlation between $V_{CME}$, $F_{SXR}$, and $I_{P}$ has therefore often been put forward \citep[e.g.][]{2016JSWSC...6A..42P}. Nonetheless, previous findings have solely been derived from an E$>$ 10 MeV sample of SEPs. Although it was neglected in most studies, we further investigate solar scaling relations here for the integrated E$>$30 Mev, E$>$ 60 Mev, and E$>$ 100 MeV energy channels based on the same method. As a result, the scaling relations $I_{P} \propto V_{CME}^{\gamma}$ are obtained for each of these integral energies (see Appendix \ref{appendix:A}). Our results are presented in Fig.~\ref{fig:fig3}, where we display the $I_P$ -- $V_{CME}$ relations for E$>$30 MeV (top panel), E$>$ 60 MeV (middle panel), and E$>$ 100 MeV (bottom panel). Table~\ref{tab:table1} summarizes the slopes obtained with the RMA regression fits for each case.

The power-law index $\gamma$ presents a relatively slight increase with energy and varies between 5.09 $\pm$ 0.78 at E$>$10 MeV up to 5.52 $\pm$ 1.02 at E$>$100 MeV, but also shows a consequent slight increase in the uncertainties. At the same time, the correlation coefficients of the $I_P$ -- $V_{CME}$ relation seem to decrease with energy.

\begin{table}[!t]
\centering
\caption[]{Slopes of the relations obtained for the peak proton flux ($I_{P}$) to the CME speed ($V_{CME}$), and the corresponding correlation coefficients for each integral energy.}
\label{tab:table1}
\begin{tabular}{lcc}
\hline
\bf{Integral Energy} & \bf{Slope $I_{P}$-$V_{CME}$} &  \bf{Correlation } \\
  \bf{(MeV)} & ($\gamma$) &   \bf{coefficient  (cc)} \\
\hline
E \textgreater{} 10       &  5.09$\pm$0.78  & 0.58\\
E \textgreater{} 30       &  5.24$\pm$0.84 & 0.54\\
E \textgreater{} 60       &  5.35$\pm$0.94 &0.44\\
E \textgreater{} 100      & 5.52$\pm$1.02 & 0.38\\
\hline
\end{tabular}%
\end{table}


\subsection{Establishing the relations}
According to \cite{2016ApJ...833L...8T}, who followed the argumentation by \cite{2012ApJ...759...71E} that the CME mass is the sum of the mass within gravitationally stratified AR corona, the kinetic energy of CMEs is proportional to the flare energy ($E_{CME} \propto E_{flare}$) and the energetic proton flux in response to the SXR class of flares can be estimated under the assumption that $F_{SXR}$ is roughly proportional to the total energy released during flares, that is, $F_{SXR} \propto E_{flare}$ (consistent with the observational findings in \cite{2012ApJ...759...71E}). $V_{CME}$ then scales with $F_{SXR}$ as
\begin{equation} \label{eq4}
 V_{CME, upper} = V_{0} \cdot F_{SXR}^{1/6} (km/s),  
\end{equation}
with $F_{SXR}$ normalized in units of 1 W/m$^{2}$ and $V_{0} = 1.3 \cdot 10^{4}$ km/s, as derived from our sample. Equation~(\ref{eq4}) is the dashed black line in Fig. \ref{fig:fig2}.

Moreover, as outlined in \cite{2016ApJ...833L...8T} and above, when we assume that the total kinetic energy of protons in SEP events ($E_{P}$) is proportional to the flare energy ($E_{flare}$) and that the duration of the proton flux enhancement is determined  by the CME propagation timescale $t_{CME}$, it follows that $E_{P} \propto I_{P} \cdot t_{CME} \propto E_{flare}$. Therefore, $I_{P}$ is scaled with $V_{CME}$ as
\begin{equation} \label{eqn}
 I_{P, upper} \propto V_{CME}^{5}.  
\end{equation}
The relations for each integral energy are positioned to run through the strongest SEP events of our sample, so that we can discuss the upper limits of the CME velocity ($V_{CME, upper}$) and peak proton flux ($I_{P, upper}$) in each plot of Figs.~\ref{fig:fig1} and \ref{fig:fig3}. 


\section{Estimating the coronal mass ejection speed, peak proton flux, and fluence for extreme solar events} \label{sec:4}

\subsection{Upper limits}
\subsubsection{Estimating $V_{CME, upper}$}

\begin{figure}[!t]
\centering
\includegraphics[width=0.48\textwidth]
{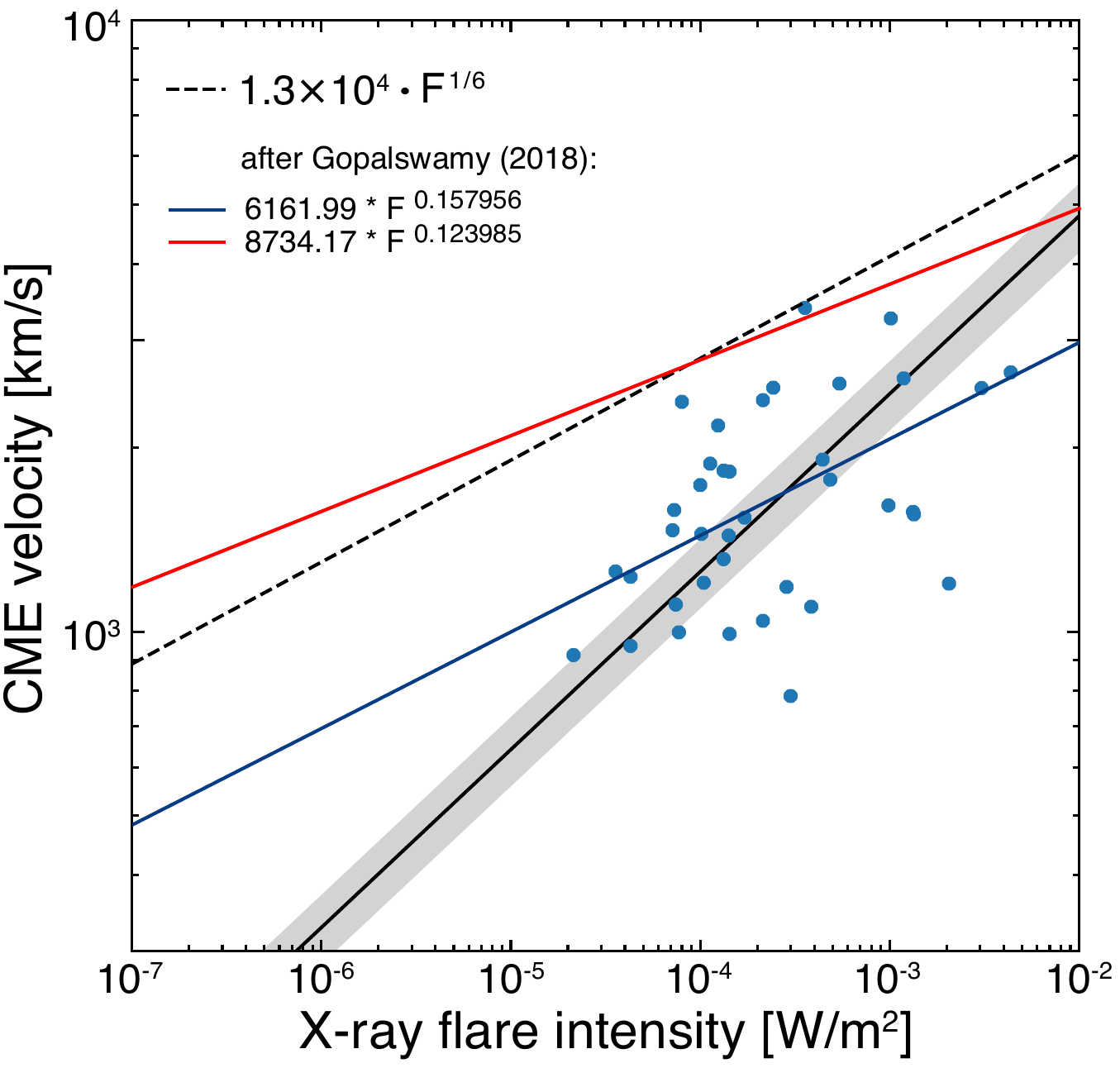}
\caption{Similar to Fig.~\ref{fig:fig2} with the addition of Eq.~(\ref{eq11}) (blue line) and Eq.~(\ref{eq12}) (red line).} 
\label{fig:fig3cme}
\end{figure}

Figure \ref{fig:fig3cme} is similar to Fig.~\ref{fig:fig2} and shows a scatter plot between $V_{CME}$ and $F_{SXR}$. The RMA fit (solid black line) is embedded into the gray error band and the upper limit is deduced from Eq.~(\ref{eq4}) and presented as the dashed black line. We investigated the upper limit of the $V_{CME}$ based on the SXR peak flux of the associated solar flare together with Eq.~(\ref{eq4}). Considering the $F_{SXR}$ values estimated by 
 \citet{2022LRSP...19....2C} of X400$\pm$200 (i.e., $4\times10^{-2}$ W/m$^2 \pm 2\times 10^{-2}$ W/m$^2$) for the AD774/775 SEP event and X28$\pm$14 (i.e., 2.8$\times 10^{-3}$ W/m$^2 \pm 1.4\times 10^{-3}$ W/m$^2$) for GLE05, we obtained $V_{CME, upper}$ =$\sim$ 7600 $\substack{+534 \\ -827}$ 
km/s and  $V_{CME, upper}$ =$\sim$ 4880$\substack{+342 \\ -532}$ 
km/s for GLE05. 
(for more details about the $F_{SXR}$ ranges, 
see \citet{2022LRSP...19....2C} and \citet{2023A&A...671A..66P}).

In the next step, known published scaling relations of $V_{CME}$ were investigated and compared with our findings. \cite{gopalswamy2018extreme} showed that the highest expected CME speed might be as high as $\sim$ 7200 km/s. This estimate was based on extreme solar conditions, which differ from those of the current Sun. Interestingly, Fig.~7(a) of \cite{gopalswamy2018extreme} provides an estimation of $V_{CME}$ and an upper limit ($V_{CME, upper}$) based on the magnetic potential energy (MPE [erg]) of active regions (ARs),  giving the corresponding empirical relations as
 \begin{equation}\label{eq9}
    V_{CME} = 748 \cdot \rm log(MPE) + 636  (km/s),
 \end{equation} 
\noindent which represents the whole sample of CMEs and ARs considered by \cite{gopalswamy2018extreme} and  
 \begin{equation} \label{eq10}
    V_{CME, upper} = 1136 \cdot \rm log(MPE) + 1557  (km/s),
 \end{equation} 
\noindent which is extracted from the highest values of the Goplaswamy sample, leading to the upper limit $V_{CME, upper}$.

According to \cite{2012ApJ...759...71E}, the bolometric flare energy is related to MPE by $F_{TSI} = 0.025 \cdot MPE$, or vice versa, $MPE = 40 \cdot F_{TSI}$. Using our 38 SEP events (see Appendix C of part I), we first substituted $F_{SXR}$ in Eq.~(1) of \cite{2020ApJ...903...41C}, which translated $F_{SXR}$ to $F_{TSI}$ for each flare. From the above relation, the MPE was estimated. 
After they were calculated for each case, the expected CME speeds and their upper limit in our sample were derived using Eqs.~(\ref{eq9}) and (\ref{eq10}). The obtained data were subsequently used to obtain fits of $V_{CME}$ ($V_{CME, upper}$) versus $F_{SXR}$, which were then added to Fig.~\ref{fig:fig2}, and we obtained the representation in Fig~\ref{fig:fig3cme}. Thereby, the obtained fits are given by
 \begin{equation} \label{eq11}
    V_{CME} = 6162 \cdot F_{SXR}^{0.158} \,\,\,\left[\sim 6.2 \times 10^{3} \cdot F_{SXR}^{1/6}\right] (km/s),
    \end{equation} 
\noindent with $F_{SXR}$ normalized in units of 1 W/m$^{2}$ and $6.2 \times 10^{3}$ in km/s. This relation provides the $V_{CME}$ $\propto$ $F_{SXR}$ relation presented as a blue line in Fig.~\ref{fig:fig3cme} and 
 \begin{equation} \label{eq12}
    V_{CME, upper} = 8734 \cdot F_{SXR}^{0.124} \,\,\,\left[\sim 8.7 \times 10^{3} \cdot F_{SXR}^{1/8}\right] (km/s),
 \end{equation} 
\noindent with $F_{SXR}$ normalized in units of 1 W/m$^{2}$ and $8.7 \times 10^{3}$ in km/s. This relation provides the $V_{CME, upper}$ $\propto$ $F_{SXR}$ relation presented as a red line in Fig.~\ref{fig:fig3cme}.
%

Table \ref{tab:cmeall} shows that $V_{CME, upper}$ for the extreme cases of AD774/775 and GLE05, derived from Eq.~(\ref{eq4}) (Fig.\ref{fig:fig3cme}; dashed black line), leads to CME speeds that are higher by $\sim$1.13-1.32 times than those obtained by Eq.~(\ref{eq12}) (solid red line in Fig.\ref{fig:fig3cme}). Nonetheless, Fig.~\ref{fig:fig3cme} shows that Eq.~(\ref{eq12}) includes all 38 SEP events of our sample, except for the event on 10 November 2004, and it provides an alternative upper limit for the CME speed compared to that proposed by \cite{2016ApJ...833L...8T} (i.e., the dashed black line in Fig.~\ref{fig:fig3cme}). Because of the difference of the obtained slopes (i.e., red versus dashed black line), the difference is clearly larger for stronger, that is, extreme flares. In other words, the stronger the flare in terms of $F_{SXR}$, the larger the difference for the different estimates of $V_{CME, upper}$. At the same time, Eq.~(\ref{eq11}) (solid blue line in Fig.\ref{fig:fig3cme}) has a similar slope as Eq.~(\ref{eq4}), but with a lower scaling. Thus, the blue line underestimates the CME speed, especially for M- and X-class flares of our sample, but provides a statistically deduced representation of the expected $V_{CME}$, in agreement with Fig.~7(a) of \cite{gopalswamy2018extreme}.  

\begin{table}[t!]
\centering
\caption[]{Upper limit CME speeds ($V_{CME, upper}$, [km/s]) for the SEP event on AD774/775 and GLE05 derived in this work (i.e., from Eq.~(\ref{eq4}) and Eq.~(\ref{eq12})), while Eq.~(\ref{eq11}) provides the $V_{CME}$ for each event.}
\label{tab:cmeall}
\begin{tabular}{lccl}
\hline
             & \bf{AD774/775} & \bf{GLE05} & \bf{Fig.3}\\
             \hline
\bf{Equation} & \multicolumn{2}{c}{\bf{CME speed -- $V_{CME}$}} & \bf{line}\\
  \bf{} & \multicolumn{2}{c}{\bf{(km/s)}} & \bf{}\\
\hline
\\[-0.5em]
Eq.~(\ref{eq4})        & 7602$_{6773}^{8134}$ & 4881$_{4348}^{5222}$ & dashed\\
\\[-0.5em]
Eq.~(\ref{eq12})       & 5860$_{5377}^{6162}$ & 4214$_{3867}^{4431}$ & red\\
\\[-0.5em]
\hline
Eq.~(\ref{eq11})      & 3706$_{3322}^{3951}$ & 2435$_{2182}^{2596}$ & blue\\
\\[-0.5em]
\hline
\end{tabular}%
\end{table}


\subsubsection{Peak fluxes and fluences of solar energetic particles driven by $V_{CME}$}

We investigated the relation between $I_{P}$, $F_{P}$, and $V_{CME}$ further. In particular, the upper limit peak proton flux  was calculated using the $I_{P}$ = $V_{energy} \cdot V_{CME}^{5}$ relations (where $V_{energy}$ are the coefficients used in the dashed black lines in Fig.~\ref{fig:fig1} and Fig. \ref{fig:fig3} for all integral SEP energies employed in this work). Similar to part I, from the established upper-limit relation of the peak proton flux, $I_{P, upper}$, the corresponding upper limit of the fluence $F_{P,upper}$ can also be retrieved. In this case, as a function of $V_{CME}$ as
%
\begin{equation} \label{eq6}
 F_{P, upper} = F_{P, energy} \cdot (V_{energy} \cdot V_{CME}^{5})^{\delta}  , 
\end{equation}
\noindent with $F_{P, energy}$ and $\delta$ directly taken from Table A.1 of \citet{2023A&A...671A..66P}\footnote{\url{https://www.aanda.org/articles/aa/full_html/2023/03/aa43407-22/T4.html}} and $V_{energy}$ being  $V_{E10}$ = 10$^{-12.05}$, $V_{E30}$ = 10$^{-12.55}$, $V_{E60}$ = 10$^{-13.14}$, and $V_{E100}$ = 10$^{-13.65}$. $V_{energy}$ for each integral energy we considered was scaled with $(km/s)^{-5}$.

The corresponding outputs are presented in Appendix \ref{appendix:B}. Figure \ref{fig:fig9} shows the $F_{P}$ versus $V_{CME}$ relation obtained for our sample of 38 SEP events for which CME information was available, using the RMA regression fit (solid black lines) for each integral energy embedded in a gray error-envelope. The dashed black lines correspond to the upper limits from Eq.~(\ref{eq6}).
\begin{figure*}[!t]
\centering
\includegraphics[width=0.9\textwidth]{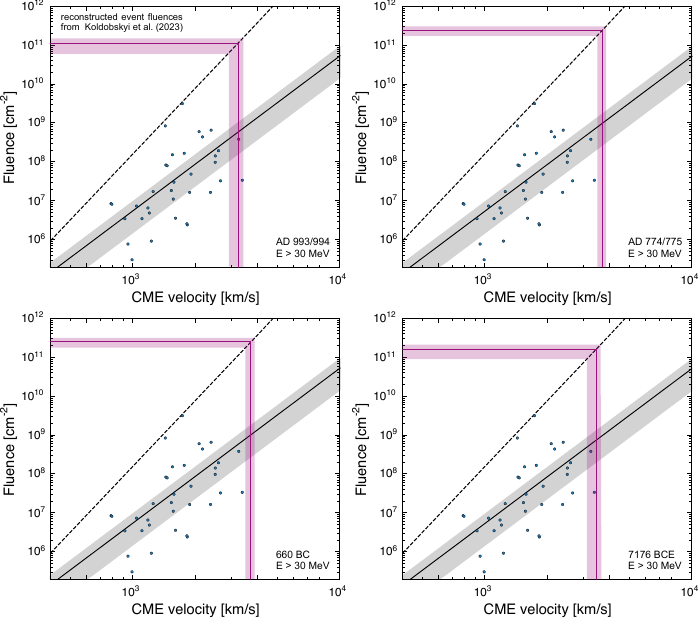}
\caption{$F_{P}$-$V_{CME}$ relations for E$>$30 MeV. 
The magenta ribbons on the Y-axis correspond to the $F_{P}$ range for AD993, AD774/775, 7176 BCE, and 660 BCE (in the clockwise direction) as published by \cite{2023JGRA..12831186K}. The ribbons on the X-axis show the estimated $V_{CME}$ range for the events based on the RMA fit (solid black line) and the upper limit (worst-case scenario; dashed black lines).}
\label{fig:fig0_2}
\end{figure*}

In order to investigate the effect of $V_{CME, upper}$ and $V_{CME}$ on the calculation of $I_{P}$ and $F_{P}$, we used the outputs from Eq.~(\ref{eq4}), Eq.~(\ref{eq12}), and Eq.~(\ref{eq11}) (see Table \ref{tab:cmeall}). Using these values (including upper and lower limits), the peak proton flux was derived employing the $I_{P} \propto V_{CME}^{5}$ relations (Figs.~\ref{fig:fig1} and \ref{fig:fig3}), and the fluence was calculated by substituting $V_{CME}$ (or $V_{CME, upper}$) in Eq.~(\ref{eq6}). 

The obtained results are provided for the upper-limit peak proton flux and fluence in Tab.~\ref{tab:res2}, showing that the upper limit $I_{P}$, which was calculated with $V_{CME, upper}$ from Eq.~(\ref{eq4}) and Eq.~(\ref{eq12}), differs by a factor of $\sim$2.5. The corresponding $F_{P, upper}$ differs by a factor of $\sim$3.0. At the same time, the obtained $I_{P}$ and $F_{P}$ are lower than the upper limits by a factor of $\sim$40 when $V_{CME}$ from Eq.~(\ref{eq11}) was used as input.  
\begin{figure*}[!t]
\centering
\includegraphics[width=0.8\textwidth]{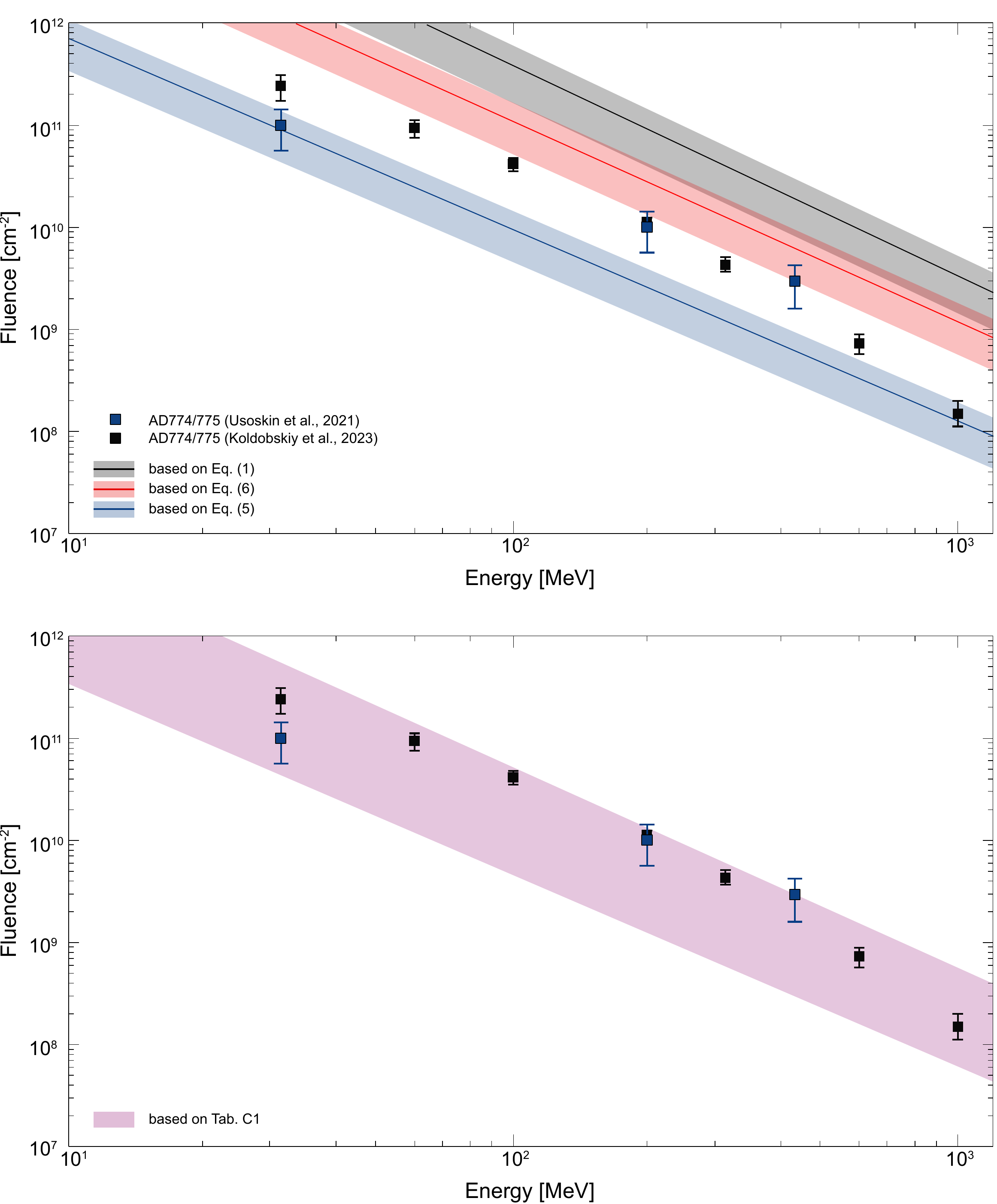}
\caption{Fluence spectra of SEPs obtained from the scaling relation $F_{P}$-$V_{CME}$ for AD774/775. Each solid line is the inverse power-law fit to the obtained integral fluence values acquired when using $V_{CME}$ from Table \ref{tab:cmeall} (top panel) and the range of CME speeds obtained from Table \ref{tab:c1} (bottom panel). The derived AD774/775 fluences by \cite{usoskin2021strongest} and \cite{2023JGRA..12831186K} are shown as filled blue and black squares, respectively.}
\label{fig:5new}
\end{figure*}
\subsection{$V_{CME}$ and its upper limit}

The upper limit of $V_{CME}$ obtained from Eq.~(\ref{eq4}) leads to a maximum value of $\sim$7600 km/s for an X425 SXR flare associated with the AD774/775 SEP event. This CME speed agrees with the value of $\sim$7200 km/s reported by \cite{gopalswamy2011coronal}. However, this high CME speed requires extreme conditions (i.e., B=6100 G over the entire sunspot region) that are difficult to reconcile with the current Sun. Moreover, the RMA fit for the same case (see the solid black line in Fig.~\ref{fig:fig3}) leads to values of $\sim$12000 km/s (for the same SXR flare of X425), which is even more difficult to present with the inherent limitations imposed by the energetics of our Sun, and thus is unrealistic. Moreover, it was shown that $V_{CME}$ from Eq.~(\ref{eq12}) provides maximum values for the SXR flare associated with the AD774/775 event of $~$5300-6100 km/s, but this decreases to $\sim$ 3300-3900 km/s when using Eq.~(\ref{eq11}) (see Table \ref{tab:cmeall} and Fig. \ref{fig:fig3cme}). 

During the modern era of CME measurements, the Sun produced an SEP event on 4 November 2003, which was associated with an X28 (X43.2\footnote{see the recent recalibration in \cite{2024SoPh..299...39H}}) solar flare and a CME with a linear speed of 2657 km/s\footnote{\url{https://cdaw.gsfc.nasa.gov/CME_list/sepe/}}. However, to determine the CME speed of this event, only three points were used in the LASCO field of view. Additionally, the CME emerged under distorted conditions, which complicates the speed estimates. Moreover, \cite{gopalswamy2005coronal} suggested that the maximum $V_{CME}$ of our host star may not be much higher than $\sim$3000 km/s.   

Figure \ref{fig:fig0_2} shows our results for the $F_{P}$ - $V_{CME}$ dependence for E$>$30 MeV. In each panel, we added the fluence of the extreme and rare SEP events found in the cosmogenic radionuclide records \citep[e.g., ][]{miyake2012signature, mekhaldi2015multiradionuclide, Brehm-etal-2021, MekhaldiEA2021}: the AD993 (upper left), AD774/775 (upper right), 660 BCE (lower left), and  7176 BCE (lower right). For these selected events, we used the recent fluence reconstructions reported by \cite{2023JGRA..12831186K}. Thus far, when we used the obtained fluence range for each of these events (Y-axis in each panel), the $V_{CME}$ range leading to this fluence can be directly obtained based on the upper limit scaling law (dashed black line). Similar figures were constructed for E$>$60 MeV (Fig. \ref{fig:figc1}) and E$>$100 MeV (Fig. \ref{fig:figc2}). When the upper limit is used, the range of the CME speed for these extreme SEPs falls within a mid-mean range of 3241 - 5255 km/s based on all integral energies (i.e., see details in Appendix \ref{appendix:C} and Table \ref{tab:c1}).

\subsection{Spectrum based on $V_{CME}$}

The top panel of Fig. \ref{fig:5new} provides the obtained integral fluence spectra of the AD774/775 event for the three different estimates of the CME speed based on the $F_{SXR}$ (see Table \ref{tab:cmeall}). In particular, the upper limit of $V_{CME}$ was obtained by Eq.~(\ref{eq4}) (dashed black line in Fig.~\ref{fig:fig3cme}), Eq.~(\ref{eq11}) (solid blue line in Fig.~\ref{fig:fig3cme}), and Eq.~(\ref{eq12}) (solid red line in Fig.~\ref{fig:fig3cme}). Thereafter, the $F_{P}$ - $V_{CME}$ relations (i.e., RMA fits in Fig.\ref{fig:fig9} and Eq.~(\ref{eq6}) for the upper limits) were employed to identify the expected fluence at the respective integral energies. The solid black, blue, and red lines were obtained via the inverse power-law fit of the estimated fluences per integral energy of the uppermost points included in Table \ref{tab:res2}. These points were calculated using Eq.~(\ref{eq6}). For each of the three fits, the shaded area provides the 1$\sigma$ error. Estimates of the energy-dependent fluences of the AD774/775 event, independently obtained by \cite{usoskin2021strongest} and \cite{2023JGRA..12831186K}, are included as blue and black squares, respectively.

The obtained integral fluence spectra displayed in the top panel of Fig.~\ref{fig:5new} are driven by the associated $V_{CME}$. The reconstructions by \cite{usoskin2021strongest} and \cite{2023JGRA..12831186K} seem to fall above the estimated fluence spectra obtained from an X425$\pm$175 flare converted into $V_{CME}$ using Eq.~(\ref{eq11}) (i.e., solid blue line) and below the upper limit ($V_{CME}^{5}$) spectra presented in this work (i.e., solid black and red lines) when converting the same flare magnitude using Eq.~(\ref{eq4}) and Eq.~(\ref{eq12}), respectively. Therefore, the obtained upper limit (i.e., the worst case) fluence spectra in these two later cases (solid black and red lines, respectively) seem to restrict the actual fluence-reconstructed values in the case of AD774/775. Nonetheless, for higher energies (i.e., E$>$430 MeV and E$>$1000 MeV), the difference between the actual estimates of the fluence (black squares) and the obtained fluence spectra (solid black line) exceeds one order of magnitude, which suggests that a CME speed of 7602 $\substack{+532 \\ -829}$ km/s overestimates the fluence at these energies.

The bottom panel of Fig. \ref{fig:5new} is similar to the top panel. However, the obtained fluence spectrum here assumes a $V_{CME}$ range of 3421 km/s (lower limit) to 5482 km/s (upper limit) as obtained in Appendix \ref{appendix:C} (magenta band). It shows that the independently obtained fluence estimates of \cite{usoskin2021strongest} and \cite{2023JGRA..12831186K} fall well within the estimated fluence range for energies between E$>$30MeV and E$>$1 GeV. When we assume that the AD774/775 event is one of the strongest events ever recorded (based on our current knowledge), the fluence spectra depicted in the bottom panel of Fig.~ \ref{fig:5new} offer an upper limit (worst case) fluence, including the independently obtained fluence values, based on a CME speed of 5482 km/s (see Appendix \ref{appendix:C} for more details). Thus, in turn, a CME speed of $\leq$5500 km/s could be restrictive for our current Sun.

\section{Conclusions}
\label{sec:conclusion}
We investigated the dependence of SEP events on $V_{CME}$. In particular, the scaling relations that describe $I_{P}$, $V_{CME}$, and $F_{P}$ seem to be consistent with statistical relations obtained by observations. These were extended from E$>$10 MeV up to E$>$100 MeV. Based on our analysis, we derived the maximum expected CME speed associated with the largest estimated $F_{SXR}$ flare that was unleashed by the Sun (i.e., X600) to range between $\sim$3950 and $\sim$8134 km/s, depending on the underlying relation (see Table \ref{tab:cmeall}). However, a CME speed as high as $\sim$8200 km/s could only result from exceptional and most likely unrealistic conditions \citep[see][]{gopalswamy2018extreme}. 

Moreover, the use of Eq.~(\ref{eq11}) showed that a CME speed of $\sim$3950 km/s agrees relatively well with CME observations of the extreme case of 4 November 2001 ($V_{CME}$=2660 km/s). This $V_{CME}$ provides an upper-limit fluence spectrum that is consistent with observations for E$>$30 MeV (blue line in Fig.~\ref{fig:5new}, top panel). However, at higher energies from E$>$60MeV to E$>$430 MeV, this CME speed leads to an underestimation of the fluence. At the same time,  the use of Eq.~(\ref{eq12}) showed that a CME speed of $\sim$5380 km/s provides an upper-limit spectrum comparable with higher energies (i.e., E$>$100 MeV, E$>$200 MeV \& E$>$430 MeV), but it overestimates the respective fluence. Nonetheless, at E$>$30 MeV, an upper-limit estimation of the fluence for AD774/775 based on $V_{CME}$ $\leq$4000 km/s seems to represent the obtained fluence better \citep{usoskin2021strongest}. 

The scaling relations presented in this study provide a direct estimate of the upper-limit peak flux ($I_{P}$) and fluence ($F_{P}$) based on the $V_{CME}^5$ relation driven by the associated CME speed. The correlation coefficients of $F_{SXR}$ and $V_{CME}$ are roughly similar in strength \citep[see][and this work]{2023A&A...671A..66P}. However, $F_{SXR}$ possibly provides more robust estimates of $I_{P}$ and $F_{P}$ than the CME speed because 
(a) the CME speed measurement is more uncertain and may be strongly variable during the early phase close to Sun \citep{2001ApJ...559..452Z,vrvsnak2008processes}, with peak accelerations as high as 10 $km/s^{2}$ \citep{2011ApJ...738..191B,2018ApJ...868..107V} , whereas the flare peak SXR flux ($F_{SXR}$) is well measured \footnote{The main reason is the projection effects, which depend on the location of the source region on the Sun \citep{paouris2021assessing}.}, and (b) the conversion of the $F_{SXR}$ into the $V_{CME}$ (Fig.~\ref{fig:fig3cme}) leads to three CME speed estimates for the same flare, depending on which relation is considered. Based on this and the scaling relations presented in \citep[part I, ][]{2023A&A...671A..66P}, $F_{SXR}$ scales almost linearly ($\gamma$ = 5/6) with $I_{P}$ and $F_{P}$, whereas $V_{CME}$ scales with a $\gamma$ = 5. As a result, the errors in determining the CME speed are strongly enhanced pertaining to the dependent quantity upon calculations. This spread of values obtained for fluences driven by the same CME speed (see Fig.~\ref{fig:fig0_2}) could also be explained by the apparent lack of $V_{CME}$ values that extend to $\geq$ 4000 km/s, which causes a critical gap that is not covered by (any of) the sample(s) used (in any such study). Additionally, one of the theoretical assumptions is that the relation $t_{CME} \propto L/V_{CME}$ holds true, where $L$ is the AR size. However, for completeness, we note that when a constant length $L_{0}$ is assumed instead, the scaling relation changes to $F_{P} \propto V_{CME}^7$. Although plausible for the acceleration of particles whose length scales are independent of the AR size, this scaling law would in fact mean that even slower CMEs would cause extreme fluences (orders of magnitude higher) at the respective energies employed in this work. Based on the measurements of CMEs of the last 27 years, this seems highly unlikely.
 
The top panel of Fig. \ref{fig:5new} shows that the highest expected fluence for a given $V_{CME}$ cannot (or can only slightly) overcome values demarcated by the red line. In addition, the bottom panel of Fig. \ref{fig:5new} highlights that applying the scaling relation $F_{P} \propto V_{CME}^{5}$ and assuming CME speeds between 3420 km/s and 5480 km/s leads to $F_{P}$ values for one of the largest SEP event observed to date (i.e., AD774/775) that agree well with the values calculated by \cite{usoskin2021strongest} and \cite{2023JGRA..12831186K}. 

We underline that these relations do not necessarily assume a pure solar flare or CME acceleration of SEPs. As noted in the pioneering work by \cite{2012ApJ...759...71E}, there is an interplay between X-ray and SEP emission in complex solar events, including CME generation, so that the energy release is distributed between radiation and accelerated particles. Moreover, scaling relations are inherent of caveats and limitations, as outlined in part I \citep{2023A&A...671A..66P}, but provide valuable content for the estimation of the worst-case radiation environment based on the $V_{CME}$ alone. 

Our results may  apply to other Sun-like stars. Nonetheless, for our host star, \cite{2021ApJ...917L..29L}  estimated that no more than 50\% of solar flares with a magnitude $\sim$X100 would generate CMEs caused by the strong magnetic confinement exerted on flares/eruptions from the largest ARs in terms of their overall magnetic flux. This finding clearly has consequences for the rate of SEP production. Nevertheless, as highlighted in part I and Sec.\ref{sec:intro}, scaling relations offer valuable context for a worst-case (upper limit) estimate of the radiation environment because these relations inherently assume flares to be associated with CMEs, and consequently, with SEP events. However, this one-to-one association scenario is unrealistic and not observed on the Sun because tens of thousands of flares result in only a few hundred SEP events \citep{2016JSWSC...6A..42P}. This finding was recently corroborated by \cite{2023ApJ...956...24K}, who concluded that scaling relations cannot be used directly without taking this imbalance into account. 

We further underlined the difficulties in identifying the upper CME speed and how they impact the resulting $I_{P}$ and $F_{P}$ values. In addition, caution is needed in the case of stellar CMEs because of the observational gap in the stellar regime: While numerous stellar flares have been observed in Sun-like stars, stellar CMEs might be rare and cannot be directly observed so far \citep[see details in][]{2019ApJ...877..105M,2022SerAJ.205....1L}. Nonetheless, recent studies demonstrated a different approach for identifying stellar CMEs on cool stars. This approach is based on the sudden dimmings in the extreme-ultraviolet and X-ray emission caused by the CME mass loss \citep{2021NatAs...5..697V, LoydEA2022, NotsuEA2024}, and it shows potential for future research efforts.


\begin{acknowledgements}
The authors would like to thank the anonymous referee for a critical and constructive reading of the manuscript and for valuable comments that improved the contents of the paper. AP, KH, and DL acknowledge the International Space Science Institute and the supported International Team 441: High EneRgy sOlar partICle Events Analysis (HEROIC). KH acknowledges the support of the DFG priority program SPP 1992 “Exploring the Diversity of Extrasolar Planets (HE 8392/1-1)”. 
AP and KH also acknowledge the supported International Team 464: The Role Of Solar And Stellar Energetic Particles On (Exo)Planetary Habitability (ETERNAL).  AP and DL acknowledge support from NASA/LWS project NNH19ZDA001N-LWS. DL also acknowledges support from project NNH17ZDH001N-LWS. AMV acknowledges the Austrian Science Fund (FWF): project no. 10.55776/I4555.
\end{acknowledgements}


\bibliographystyle{aa}
\bibliography{references}

\appendix

\section{The $I_{P}$ versus $V_{CME}$ relations for different integral energies} \label{appendix:A}

\begin{figure}[h!]
\centering
\includegraphics[width=0.81\columnwidth]{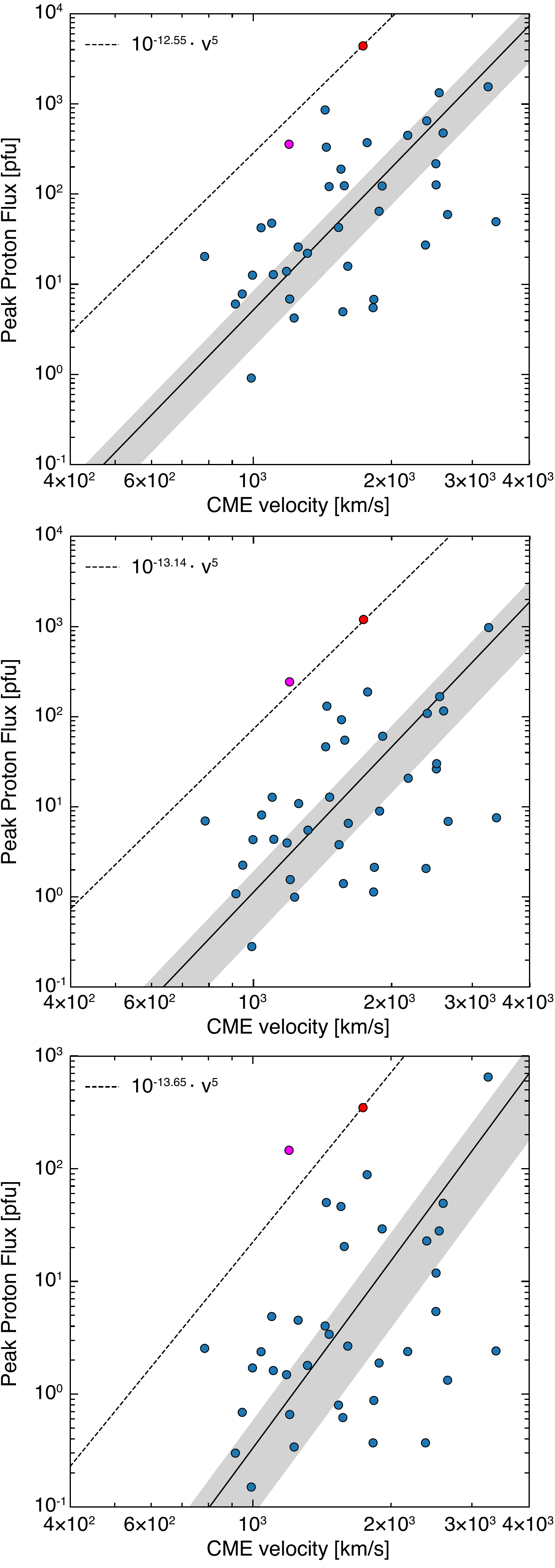}
\caption{Similar to Fig. \ref{fig:fig1}. From top to bottom, these panels present results for E$>$30-; E$>$60-, and E$>$100 MeV, respectively. In all panels, the red dot corresponds to the 8 November 2000 outstanding large SEP event (see text for further details).} 
\label{fig:fig3}
\end{figure}
Here, we present the same scatter plots as for Fig. \ref{fig:fig1} but for $I_{P}$ at energies E$>$30 MeV, E$>$60 MeV, and E$>$100 MeV. The red and magenta dots indicate the events on 8 November 2000 and 15 April 2001, respectively. Our scaling is based on the 8 November 2000 event. This is the event that achieves the highest peak proton flux across all energies considered (i.e. E$>$10 - E$>$100 MeV). From the interplay of the associated $V_{CME}$ with the achieved peak proton flux, the 8 November 2000 SEP event is clearly distinguished at E$>$10 MeV (Fig.~ \ref{fig:fig1}) and E$>$30 MeV (Fig.~ \ref{fig:fig3}, top panel). For E$>$60 MeV the 15 April 2001 SEP event falls on the upper-limit scaling deduced by the 8 November 2000 SEP event (Fig.~ \ref{fig:fig3}, dashed black line in the middle panel). However, for E$>$100 MeV, although the achieved peak proton flux on 15 April 2001 is lower than that of the 8 November 2000 SEP event (Fig.~ \ref{fig:fig3}, bottom panel; y-axis) this SEP event is associated to a CME with a speed of 1199 km/s - which is lower than the CME speed of 1738 km/s associated with the 8 November 2000 SEP event. As a result, the magenta point seems to differentiate from the upper-limit scaling (dashed black line) by a factor of $\sim$2.6. Nonetheless, for consistency and in line with part I (see Appendix B of that work) we keep the upper-limit scaling bound to the 8 November 2000 SEP event (red dot) across all energies and propagate the relative error imposed by the differentiation at E$>$100 MeV to our calculations.

\section{The $F_{P}$ versus $V_{CME}$ relations} \label{appendix:B}

\begin{figure}[!h]
\centering
\includegraphics[width=\columnwidth]{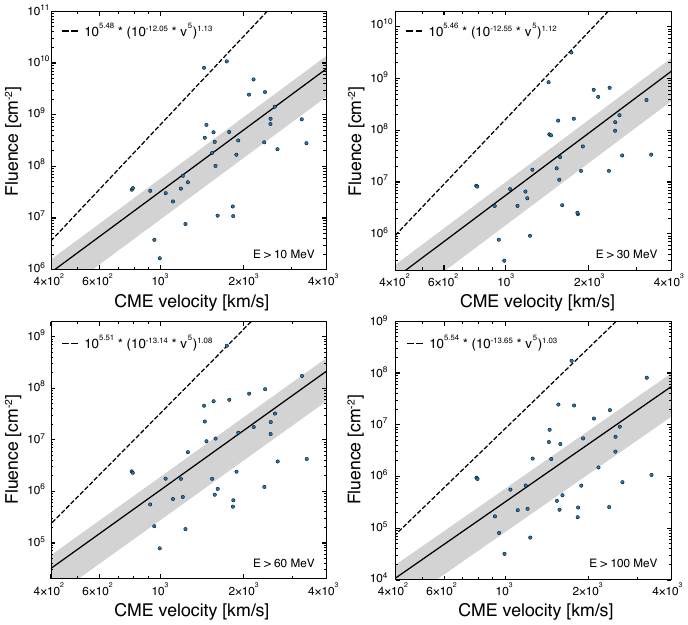}
\caption{Relation between SEP fluence ($F_{P}$) and CME speed ($V_{CME}$) for the four integral energy bands of the SEP events, i.e., E$>$10-; E$>$30-; E$>$60-; and E$>$100 MeV, respectively. The log-log relations are obtained with RMA regression fitting. The estimated upper limits of $F_{P}$ in terms of $V_{CME}$ are depicted as dashed black lines in each panel.}
\label{fig:fig9}
\end{figure}

 Figure \ref{fig:fig9} depicts the $F_{P}$-$V_{CME}$ relation obtained for our sample of 38 SEP events for which CME information was available. The RMA regression fit is presented as a solid black line at each integral energy (i.e. panel) embedded in a gray error-envelope. The dashed black lines represent the upper-limits obtained by using Eq.~(\ref{eq6}). In each case, the obtained dashed black line yields an upper-limit to the observed fluences.
 
\begin{table*}[]
\begin{center}
\caption[]{Upper limit peak proton fluxes ($I_{P}$, [pfu]) and fluences ($F_{P}$, [cm$^{-2}$]) for the SEP event on AD774/775 and GLE05 derived in this work, for each integral proton energy.}
\label{tab:res2}
\begin{tabular}{lcccccc}
\hline
& \multicolumn{2}{c}{\bf{$V_{CME, upper}$ | Eq.(\ref{eq4})}} & \multicolumn{2}{c}{\bf{$V_{CME, upper}$ | Eq.(\ref{eq12})}} & \multicolumn{2}{c}{\bf{$V_{CME}$ | Eq.(\ref{eq11})}}\\
             \hline
             & \bf{AD774/775} & \bf{GLE05} & \bf{AD774/775} & \bf{GLE05}  & \bf{AD774/775} & \bf{GLE05} \\
             \hline
             \bf{Integral} & & & & & \\
\bf{Energy} & \multicolumn{2}{c}{\bf{Peak Proton Flux - $I_{P}$}} & \multicolumn{2}{c}{\bf{Peak Proton Flux - $I_{P}$}} & \multicolumn{2}{c}{\bf{Peak Proton Flux - $I_{P}$}}\\
  \bf{(MeV)} & \bf{(pfu)} & \bf{(pfu)} & \bf{(pfu)} & \bf{(pfu)} & \bf{(pfu)} & \bf{(pfu)} \\
\hline
\\[-0.5em]
E\textgreater{}10       & 2.26E+07$_{1.27\rm E+07}^{3.17\rm E+07}$ & 2.47E+06$_{1.39\rm E+06}^{3.46\rm E+06}$ & 6.16E+06$_{4.01\rm E+06}^{7.92\rm E+06}$ & 1.18E+06$_{7.71\rm E+05}^{1.52\rm E+06}$ & 6.23E+05$_{3.60\rm E+05}^{8.58\rm E+05}$ & 7.63E+04$_{4.41\rm E+04}^{1.05\rm E+05}$\\
\\[-0.5em]
E\textgreater{}30       & 7.16E+06$_{4.02\rm E+06}^{1.00\rm E+07}$ & 7.80E+05$_{4.38\rm E+05}^{1.09\rm E+06}$ & 1.95E+06$_{1.27\rm E+06}^{2.50\rm E+06}$ & 3.75E+05$_{2.44\rm E+05}^{4.82\rm E+05}$ & 1.97E+05$_{1.14\rm E+05}^{2.71\rm E+05}$ & 2.41E+04$_{1.40\rm E+04}^{3.32\rm E+04}$\\
\\[-0.5em]
E\textgreater{}60       & 1.84E+06$_{1.03\rm E+06}^{2.58\rm E+06}$ & 2.01E+05$_{1.13\rm E+05}^{2.81\rm E+05}$ & 5.01E+05$_{3.26\rm E+05}^{6.44\rm E+05}$ & 9.63E+04$_{6.27\rm E+04}^{1.24\rm E+05}$ & 5.06E+04$_{2.93\rm E+04}^{6.98\rm E+04}$ & 6.20E+03$_{3.59\rm E+03}^{8.54\rm E+03}$\\
\\[-0.5em]
E\textgreater{}100      & 5.69E+05$_{3.19\rm E+05}^{7.97\rm E+05}$ & 6.20E+04$_{3.48\rm E+04}^{8.69\rm E+04}$ & 1.55E+05$_{1.01\rm E+05}^{1.99\rm E+05}$ & 2.98E+04$_{1.94\rm E+04}^{3.83\rm E+04}$ & 1.57E+04$_{9.05\rm E+03}^{2.16\rm E+04}$ & 1.92E+03$_{1.11\rm E+03}^{2.64\rm E+03}$\\
\\[-0.5em]
\hline
\bf{Integral} & & & & & \\
\bf{Energy} & \multicolumn{2}{c}{\bf{Fluence - $F_{P}$}} & \multicolumn{2}{c}{\bf{Fluence - $F_{P}$}} & \multicolumn{2}{c}{\bf{Fluence - $F_{P}$}} \\
  \bf{(MeV)} & \bf{(cm$^{-2}$)} & \bf{(cm$^{-2}$)} & \bf{(cm$^{-2}$)} & \bf{(cm$^{-2}$)} & \bf{(cm$^{-2}$)} & \bf{(cm$^{-2}$)}\\
\hline
\\[-0.5em]
E\textgreater{}10       & 6.18E+13$_{3.22\rm E+13}^{9.05\rm E+13}$ & 5.05E+12$_{2.63\rm E+12}^{7.40\rm E+12}$ & 1.42E+13$_{8.73\rm E+12}^{1.89\rm E+13}$ & 2.20E+12$_{1.36\rm E+12}^{2.93\rm E+12}$ & 1.07E+12$_{5.74\rm E+11}^{1.53\rm E+12}$ & 9.93E+10$_{5.35\rm E+10}^{1.43\rm E+11}$\\
\\[-0.5em]
E\textgreater{}30       & 1.37E+13$_{7.18\rm E+12}^{2.00\rm E+13}$ & 1.15E+12$_{6.00\rm E+11}^{1.67\rm E+12}$ & 3.19E+12$_{1.97\rm E+12}^{4.23\rm E+12}$ & 5.04E+11$_{3.11\rm E+11}^{6.68\rm E+11}$ & 2.45E+11$_{1.33\rm E+11}^{3.51\rm E+11}$ & 2.34E+10$_{1.26\rm E+10}^{3.34\rm E+10}$\\
\\[-0.5em]
E\textgreater{}60       & 1.89E+12$_{1.01\rm E+12}^{2.72\rm E+12}$ & 1.72E+11$_{9.24\rm E+10}^{2.48\rm E+11}$ & 4.63E+11$_{2.91\rm E+11}^{6.07\rm E+11}$ & 7.80E+10$_{4.91\rm E+10}^{1.02\rm E+11}$ & 3.90E+10$_{2.16\rm E+10}^{5.51\rm E+10}$ & 4.03E+09$_{2.23\rm E+09}^{5.70\rm E+09}$\\
\\[-0.5em]
E\textgreater{}100      & 2.93E+11$_{1.62\rm E+11}^{4.15\rm E+11}$ & 2.99E+10$_{1.65\rm E+10}^{4.24\rm E+10}$ & 7.68E+10$_{4.93\rm E+10}^{9.95\rm E+10}$ & 1.41E+10$_{9.03\rm E+09}^{1.82\rm E+10}$ & 7.25E+09$_{4.13\rm E+09}^{1.01\rm E+10}$ & 8.83E+08$_{4.74\rm E+08}^{1.16\rm E+09}$\\
\\[-0.5em]
\hline
\end{tabular}%
\tablefoot{Peak proton fluxes were calculated via the $I_{P} \propto V_{CME}^{5}$ relations (provided in Figs.~\ref{fig:fig1} \& \ref{fig:fig3}) and Fluence via Eq. \ref{eq6} for a given $V_{CME, upper}$. The upper and lower limits included in the Table are driven by the $V_{CME}$ range of the associated solar flare per event. The second and third columns provide outputs based on $V_{CME, upper}$ from Eq.~(\ref{eq4}). The fourth and fifth columns provide the same results based on $V_{CME, upper}$ from Eq.~(\ref{eq12}) and the sixth and seventh columns provide the same results based on $V_{CME}$ from Eq.~(\ref{eq11}).}
\end{center}

\end{table*}

\section{$F_P$-$V_{CME}$ relations for E $>$ 60 MeV and E $>$ 100 MeV} \label{appendix:C}

\begin{eqnarray}
    F_{\mathrm{E>30~MeV}} =& 10^{5.46} \cdot \left(10^{-12.55}\cdot V_{\mathrm{CME}}^5\right)^{1.12}\\ 
    =& 2.53513\cdot 10^{-9}\cdot V_{\mathrm{CME}}^{5.6}\nonumber\\
    V_{\mathrm{CME~(E>30~MeV)}} =& 34.2768 \cdot F_{\mathrm{E>30~MeV}}^{0.178571} \label{eq:C2}\\
    &\nonumber\\
    \hline\nonumber\\
        &\nonumber\\
    F_{\mathrm{E>60~MeV}} =& 10^{5.51} \cdot \left(10^{-13.14}\cdot V_{\mathrm{CME}}^5\right)^{1.08}\\ 
    =& 2.08353\cdot 10^{-9}\cdot V_{\mathrm{CME}}^{5.4}\nonumber\\
    V_{\mathrm{CME~(E>60~MeV)}} =& 40.5163 \cdot F_{\mathrm{E>60~MeV}}^{0.185185}\label{eq:C4}\\
        &\nonumber\\
    \hline\nonumber\\
        &\nonumber\\
    F_{\mathrm{E>100~MeV}} =& 10^{5.54} \cdot \left(10^{-13.65}\cdot v^5\right)^{1.03}\\ 
    =& 3.0234\cdot 10^{-9}\cdot v^{5.15}\nonumber\\
    V_{\mathrm{CME~(E>100~MeV)}} =& 45.11 \cdot F_{\mathrm{E>100~MeV}}^{0.194175}\label{eq:C6}
  \end{eqnarray}

\begin{table*}[]
\centering
\caption{The integral fluence $F_{P}$ of the extreme SEP events and the resulting $V_{CME}$ per integral energy.} \label{tab:c1}
\begin{tabular}{lcccccc}
\hline
                                                      & {\textbf{F(E\textgreater{}30MeV)}*} & {\textbf{v(E\textgreater{}30MeV)}**} & {\textbf{F(E\textgreater{}60MeV)}*} & {\textbf{v(E\textgreater{}60MeV)}**} & {\textbf{F(E\textgreater{}100MeV)}*} & {\textbf{v(E\textgreater{}100MeV)}**} \\
                                                      & {\textbf{[}cm$^{-2}${]}}                & {\textbf{[}km/s{]}}                                     & {[}cm$^{-2}${]}                & {\textbf{[}km/s{]}}                                     & {\textbf{[}cm$^{-2}${]}}                 & {\textbf{[}km/s{]}}                                      \\ \hline
                                                      & {1.57$\cdot 10^{11}$}                 & {3422}             & {4.90$\cdot 10^{10}$}                 & {3866}             & {2.04$\cdot 10^{10}$}                  & {4530}              \\
\multicolumn{1}{l}{{\textbf{994 CE}}}   & {1.16$\cdot 10^{11}$}                 & {3242}              & {3.90$\cdot 10^{10}$}                 & {3706}             & {1.72$\cdot 10^{10}$}                  & {4383}              \\
                                                      & {6.30$\cdot 10^{10}$}                 & {2907}             & {3.20$\cdot 10^{10}$}                 & {3573}             & {1.42$\cdot 10^{10}$}                  & {4223}              \\ \hline
                                                      & {3.10$\cdot 10^{11}$}                 & {3864}             & {1.12$\cdot 10^{11}$}                 & {4506}             & {4.76$\cdot 10^{10}$}                  & {5341}              \\
\multicolumn{1}{l}{{\textbf{775 CE}}}   & {2.42$\cdot 10^{11}$}                 & {3697}             & {9.40$\cdot 10^{10}$}                 & {4362}             & {4.13$\cdot 10^{10}$}                  & {5195}               \\
                                                      & {1.73$\cdot 10^{11}$}                 & {3482}             & {7.50$\cdot 10^{10}$}                 & {4183}             & {3.52$\cdot 10^{10}$}                  & {5037}              \\ \hline
                                                      & {4.06$\cdot 10^{11}$}                 & {4054}             & {1.36$\cdot 10^{11}$}                 & {4671}              & {5.45$\cdot 10^{10}$}                  & {5483}              \\
\multicolumn{1}{l}{{\textbf{660 BCE}}}  & {2.41$\cdot 10^{11}$}                 & {3694}             & {9.90$\cdot 10^{10}$}                 & {4404}             & {4.38$\cdot 10^{10}$}                  & {5255}              \\
                                                      & {1.34$\cdot 10^{11}$}                 & {3326}             & {6.80$\cdot 10^{10}$}                 & {4108}             & {3.41$\cdot 10^{10}$}                  & {5006}              \\ \hline
                                                      & {2.12$\cdot 10^{11}$}                 & {3610}              & {9.50$\cdot 10^{10}$}                 & {4371}             & {4.89$\cdot 10^{10}$}                  & {5369}              \\
\multicolumn{1}{l}{{\textbf{7176 BCE}}} & {1.62$\cdot 10^{11}$}                 & {3441}             & {7.80$\cdot 10^{10}$}                 & {4214}             & {4.01$\cdot 10^{10}$}                  & {5166}              \\
                                                      & {9.00$\cdot 10^{10}$}                 & {3098}             & {4.30$\cdot 10^{10}$}                 & {3774}             & {2.37$\cdot 10^{10}$}                  & {4664}              \\ \hline
\end{tabular}
\tablefoot{Column 1 gives the extreme SEP events. Columns 2, 4 \& 6 provide the integral fluence $F_{P}$ values for E$>30$ MeV, E$>$60 MeV and E$>$100 MeV taken from \citet{2023JGRA..12831186K}. Those correspond to a mean value with an upper and lower limit per event. These columns are marked with an (*). Columns 3, 5 \& 7 represented the derived $V_{CME}$ directly obtained from Eqs. \ref{eq:C2}, \ref{eq:C4}, and \ref{eq:C6}, per integral energy, respectively. The range of values considering all the integral fluence values $F_{P}$ are: lower limit range: 2907 km/s - 5037 km/s, mean range: 3242 km/s - 5255 km/s; upper limit range: 3422 km/s - 5483 km/s.}
\end{table*}

\begin{figure*}
\centering
\includegraphics[width=0.9\textwidth]{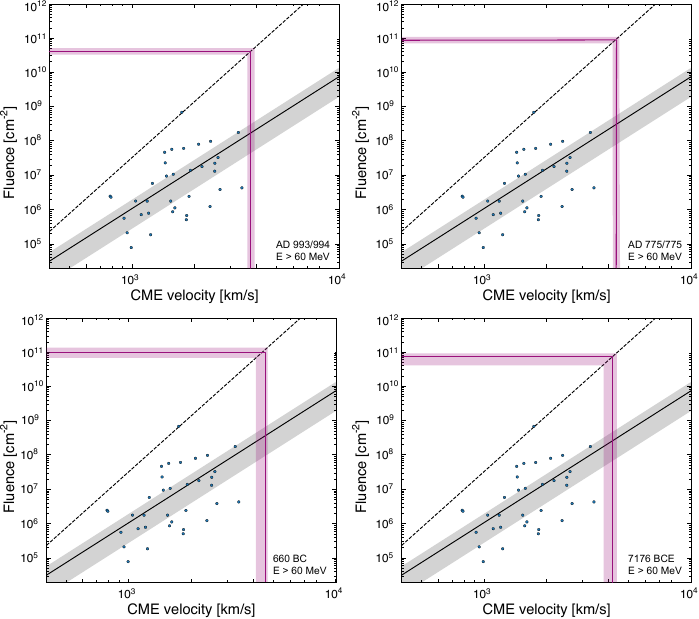}
\caption{Same as Fig.~\ref{fig:fig0_2} but for E$>$60 MeV.}
\label{fig:figc1}
\end{figure*}

\begin{figure*}
\centering
\includegraphics[width=0.9\textwidth]{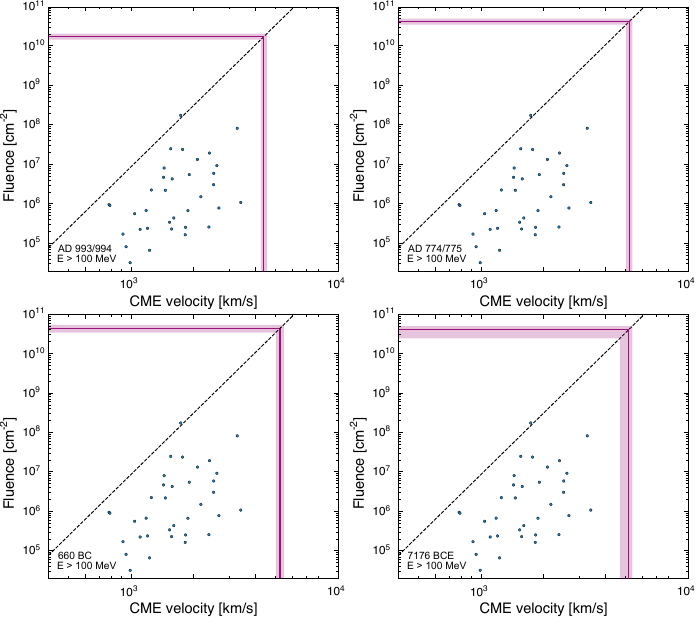}
\caption{Same as Figs.~\ref{fig:fig0_2} and ~\ref{fig:figc1}  but for E$>$100 MeV.}
\label{fig:figc2}
\end{figure*}
\end{document}